\documentstyle[12pt,aaspp4]{article}
\begin{document}
\singlespace

\title{THE STELLAR CONTENT NEAR THE GALACTIC CENTER}

\author{\bf T. J. Davidge}
\affil{Canadian Gemini Project Office, Herzberg Institute of Astrophysics, 
National Research Council of Canada, 5071 W. Saanich Road, Victoria, BC, 
Canada V8X 4M6 $^a$\\ and\\
Department of Physics \& Astronomy, University of British Columbia,
Vancouver, BC Canada V6T 1Z4\\
{\it email: tim.davidge@hia.nrc.ca}}

\author{\bf D. A. Simons} 
\affil{Gemini Project Office, PO Box 26732, 950 N. Cherry Ave, 
Tucson, Arizona, USA 85726-6732\\
{\it email: dsimons@gemini.edu}}

\author{\bf F. Rigaut}
\affil{Canada-France-Hawaii Telescope, PO Box 1597, 
Kamuela, HI USA 96743\\
{\it email: rigaut@cfht.hawaii.edu}}

\author{\bf R. Doyon}
\affil{D\'{e}partement de Physique, Universit\'{e} de Montr\'{e}al, 
C.P. 6128, Succ. Centre Ville, Montreal, PQ Canada H3C 3J7\\
{\it email: doyon@astro.umontreal.ca}}

\author{\bf D. Crampton}
\affil{ Dominion Astrophysical Observatory, Herzberg Institute of 
Astrophysics, National Research Council of Canada, 5071 W. Saanich Road, 
Victoria, BC, Canada V8X 4M6\\
{\it email: david.crampton@hia.nrc.ca}}

\slugcomment{\it To appear in the Astronomical Journal}

\noindent$^a$ Postal address

\clearpage

\begin{abstract}

	High angular resolution $J, H, K,$ and $L'$ images are used to 
investigate the stellar content within 6 arcsec of SgrA*. 
The data, which are complete to $K \sim 16$, are the deepest 
multicolor observations of this region published to date. The mean 
locus of the $(K, H-K)$ CMD varies across the field, a result that is 
attributed to differential reddening with amplitude $\Delta$A$_V \sim 15$ mag. 
The reddening variations within 3 arcsec of SgrA* are significantly smaller 
than this, and the resolved members of the compact star cluster 
immediately surrounding SgrA* have photometric properties 
that are not significantly different from objects at larger radii. 
We find that sources in our field with published $2\mu$m spectra 
showing either line emission or CO absorption occupy different sequences 
on the $(K, J-K)$ CMD. The emission line stars, which fall along the most 
richly populated sequence, have redder $J-K$ colors than 
stars in the Magellanic Clouds with similar spectroscopic characteristics, and 
evidence is presented that this is due to excess infrared 
emission in the spectrum of the GC sources. 

	The photometric properties of the giant branch, which is defined 
by stars showing CO absorption, are similar to those of the giant branch in 
Baade's Window (BW). The mean $J-K$ color and peak $K$ 
brightness of the red giant branch are both consistent with a 
metal-rich population having an age $\sim 10$ Gyr, while the width of the 
giant branch on the $(K, J-K)$ CMD is indicative of an age spread 
$\Delta log(t) \leq 1$ dex. Therefore, if the inner 
bulge contains an underlying population of stars with ages in excess of $\sim 
10$ Gyr, as is the case in BW, then the region within a few arcsec of SgrA* 
cannot contain a large population of giant branch stars younger than 1 Gyr. 
We also report the detection of a modest population 
of faint, blue sources with $K \geq 14$. We speculate that these  
are bright main sequence stars at the distance of the GC, although spectra 
will be required to confirm this interpretation. Nevertheless, the 
photometric properties of the largely unresolved compact knot of stars 
immediately surrounding SgrA* provide indirect evidence to support the 
presence of a large population of faint blue stars. In particular, 
we confirm previous studies that measure a relatively blue color
for this cluster. 

\end{abstract}

\clearpage

\section{Introduction}

	The Galactic Center (hereafter GC) provides a unique 
laboratory for investigating the nuclear stellar content of a type Sbc (Blanco 
\& Terndrup 1990, van den Bergh 1975) spiral galaxy. 
Crowding has restricted traditional ground-based surveys of 
the area within a few arcsec of SgrA* to a relatively small number of 
bright, highly evolved objects, many of which appear to be the result of a 
recent burst of star formation (Blum {\it et al.} 1996a, and references 
therein). However, the recent development of moderately large 
format infrared detector arrays, coupled with technical advances that permit 
near diffraction-limited angular resolution from ground-based sites, 
now make it possible to study fainter stars in the area immediately 
surrounding SgrA*.

	The age distribution of stars near SgrA* 
is of fundamental interest, and a mixture of young, intermediate-age, and 
old populations is likely present. The discovery of stars showing M 
supergiant (Lebofsky, Rieke, \& Tokunaga 1982) and Wolf-Rayet spectroscopic 
characteristics (Allen, Hyland, \& Hillier 1990; Libonate {\it et al.} 1995; 
Blum, DePoy, \& Sellgren 1995) within a few parsec of SgrA* suggests that star 
formation occured as recently as 10 Myr in the past (e.g. Krabbe {\it et al.} 
1995, Tamblyn \& Rieke 1993). Nevertheless, the physical conditions near the 
GC may act to suppress star formation, leading some to suggest 
alternatative explanations for the bright blue objects detected near the GC 
(e.g. Morris 1993). 

	There is evidence that the most recent burst of 
star formation near the GC was not a unique event. In particular,
Blum, Sellgren, \& DePoy (1996b) investigated the spectroscopic properties of 
bright red stars within an arcmin of the GC, and concluded that many of these 
are evolving on the asymptotic giant branch (AGB), and formed during 
intermediate epochs. Less direct evidence comes from 
studies of the nuclear regions of M31, which 
indicate that a centrally-concentrated intermediate-age component is 
also present in that galaxy (Rich \& Mighell 1995; Davidge {\it et al.} 1997).

	Although largely unstudied, an underlying population of old stars 
should also be present near the GC. In fact, photometric 
(Holtzmann {\it et al.} 1993) and spectroscopic (Idiart, de Freitas Pacheco, 
\& Costa 1996) studies of Baade's Window (hereafter BW) 
indicate that the bulge has an age in excess of 10 Gyr. There are 
indications that the region within a few arcmin of the GC contains a 
bulge population that is closely related to that in BW. For example, the $K$ 
luminosity function (LF) of stars within an arcmin of the GC constructed by 
Blum {\it et al.} (1996a) is well matched by the BW LF when $M_K \geq -7$ after 
adjusting for the surface brightness differences between the two fields. In 
addition, spectroscopic studies of nearby galaxies (e.g. Davidge 1997) indicate 
that the bulge population extends into the nuclear regions of these objects. 
Nevertheless, the photometric properties of red giants within a few arcsec 
of SgrA* remain uncertain, although the detection of a deficiency in the 
strength of CO absorption (Haller {\it et al.} 1996; Sellgren et 
al. 1990) is ostensibly suggestive of a unique population of objects.

	One of the primary goals of population studies is to establish ages, 
a task which in turn requires the detection of the main sequence turn-off. 
For the GC, this challenging task will require 
a combination of deep photometric and spectroscopic observations 
which, with the possible exception of the youngest population, are 
currently not feasible. Nevertheless, near-diffraction limited imaging studies 
conducted with 4 metre telescopes can still provide insight into 
the stellar content near the GC. The $1 - 2.5\mu$m wavelength region is of 
particular interest for imaging studies, as it provides a good compromise 
between the needs to reduce line-of-site extinction while avoiding the high 
background levels that occur longward of $2.5\mu$m. 
Line blanketing, which can significantly affect the spectral energy 
distributions of moderately metal-rich cool stars 
at optical wavelengths (Bica, Barbuy, \& Ortolani 1991), is also greatly 
reduced longward of $1\mu$m. Finally, imaging information is 
required to conduct astrometric studies, and build upon the pioneering 
proper motion studies conducted by Eckart \& Genzel (1996). 

	Attempts to achieve near diffraction-limited angular resolutions 
near the GC have been done both during (Simons \& 
Becklin 1996; Close, McCarthy, \& Melia 1995; Eckart {\it et al.} 
1993, 1995) and after (Depoy \& Sharpe 1991) data acquisition, with the goal 
of investigating the properties of relatively bright sources. 
In the current paper, deep high angular resolution $JHK$ images 
obtained with the 3.6 metre Canada-France-Hawaii Telescope (CFHT) 
adaptive optics bonnette (AOB), are used to investigate the properties of 
stars fainter than $K = 16$ in a $12 \times 12$ arcsec field 
centered on SgrA*. These data, which Rigaut et 
al. (1997a) use to measure stellar proper motions in the vicinity of SgrA*, 
are supplemented with the deep $L'$ image constructed by 
Simons \& Becklin (1996), which has been re-analyzed for this study. 
To the best of our knowledge, this is the deepest multicolor 
survey of stellar content near SgrA* conducted to date. 

\section{Observations and Reductions}

\subsection{JHK Observations}

	The data were recorded with the CFHT Adaptive Optics Bonnette (AOB) 
and MONICA (Nadeau {\it et al.} 1994) during the nights of UT June 
21/22 and 22/23 1996. Optics were installed 
in MONICA that produced an image scale of 0.034 arcsec per pixel, so that the 
$256 \times 256$ Hg:Cd:Te array sampled an $8.7 \times 8.7$ arcsec field. 
The key elements of the AOB are a 19 element curvature wavefront 
sensor, and a 19 electrode bimorph deformable 
mirror. The optics are conjugated to the telescope pupil. The system 
operates under modal control with mode gain optimization (Gendron \& 
L\'{e}na 1994) to track variations in atmospheric seeing. A more detailed 
description of the AOB is given by Rigaut {\it et al.} (1997b). 

	A complete observing sequence in each filter consisted of six 
exposures, offset on the sky at 2 arcsec intervals such that the centers of the 
various images defined a $4 \times 6$ arcsec 
rectangle. Four observing sequences were completed in $K$, 
while a single sequence was completed in the other filters. 
The reference source for atmospheric compensation was a $V = 14$ star 10 arcsec 
East and 20 arcsec North of SgrA*. Additional details 
of the observations, including integration times and 
image quality measurements, are given in Table 1. Two estimates of image 
quality were computed for each filter: the FWHM, which measures the central 
compactness of the PSF, and the diameter of an aperture containing 50\% of the 
total energy, which measures the amount of light in the extended wings of the 
PSF. These two quantities are similar at low Strehl ratios.

	The data were reduced using the following sequence, which is similar 
to that described by Hodapp, Rayner, \& Irwin (1992): (1) dark 
subtraction; (2) division by a dome flat; (3) replacement 
of bad pixels by interpolating between adjacent pixels; (4) 
subtraction of the DC sky level, estimated for each image using the DAOPHOT 
(Stetson 1987) SKY routine; and (5) for the $K$ images only, subtraction 
of the thermal component, which is introduced by foreign material (e.g. dust) 
on, and thermal inhomogeneities within, the optical elements. The thermal 
calibration frame was constructed by median combining 
flat-fielded and sky-subtracted $K$ images of a blank sky field, which 
was observed immediately following the GC. We emphasize that the thermal 
component is additive, and should not be confused with the multiplicative 
variations that are removed during flat-fielding.

	The reduced data were combined to create two sets of images for 
subsequent analysis. A field covering $12.4 \times 11.9$ arcsec was 
mosaiced in each filter by combining selected 
portions of individual exposures. There was no 
overlap between the portions of the images that were combined, so 
the integration time at each pixel in the final mosaic 
is that of a single exposure. Frame-to-frame differences in image quality were 
found to be negligible, so it was not necessary to adjust the image quality 
in the individual frames when constructing the mosaic. The resulting images 
were used for the photometric analysis discussed in \S 4. The final $K$ 
mosaic is shown in Figure 1.

	A second set of deeper images, covering a $4.7 \times 4.7$ 
arcsec field centered on SgrA*, was constructed by averaging those 
portions of the exposures common to the six offset positions. The effective 
integration time of these images is six times that of the individual exposures. 
These deep images are used in \S 5 to investigate the integrated photometric 
properties of the star cluster immediately surrounding SgrA*. 
The final deep $K$ image is shown in Figure 2.

\subsection{$L'$ Observations}

	Details concerning the acquisition and reduction of the $L'$ data 
were presented by Simons \& Becklin (1996). The final $L'$ image from 
that study was rotated and spatially re-sampled to match the orientation and 
pixel scale of the $JHK$ data. The $L'$ image overlaps with almost 60\% 
of the field mosaiced with the AOB.

\section{Photometric Measurements}

	The brightnesses of individual stars were measured using the 
PSF-fitting routine ALLSTAR (Stetson \& Harris 1988), which is part of the 
DAOPHOT (Stetson 1987) photometry package. Between 3 and 6 bright stars were 
used to construct the PSFs, depending on the filter. Isoplanatic effects were 
found to be negligible due to the small field size. The photometric calibration 
of the CFHT data was set with the UKIRT faint standard star FS24 (Casali \& 
Hawarden 1992), which was observed at various times throughout the four night 
observing run. Casali \& Hawarden (1992) find that $K$ and $J-K$ for FS24 are 
repeatable to $\pm 0.008$ mag and $\pm 0.006$ mag, respectively. The 
instrumental $L'$ measurements were calibrated by computing the zeropoint shift 
needed to reproduce the IRS7 $L$ brightness published by Forrest, Pipher, \& 
Stein (1989). Although there are indications that IRS7 is variable, with an 
amplitude of $\pm 0.1$ mag in $K$ (e.g. Blum {\it et al.} 1996a), the $L$ 
measurements of sources in the IRS16 complex made from the current data are in 
excellent agreement with those made by Depoy \& Sharpe (1991, see below), 
suggesting that any error in the calibration is not large.

	The radially-averaged intensity and encircled energy profiles derived 
from the $J, H,$ and $K$ PSFs are shown in Figure 3. The intensity curves 
have been scaled so that the total energies under each 
curve are identical; hence, the peak intensities measure relative 
Strehl ratios. It is apparent that the Strehl ratio increases with wavelength, 
as predicted by simple models of atmospheric turbulence (e.g. Fried 1965). 
In addition, even though the central diffraction peak is 
clearly evident in the $H$ and $K$ PSFs, much of the total PSF energy is 
contained in broad wings that extend out to $0.7 - 0.8$ arcsec radii. These 
extended wings are the main reason that the two 
image quality estimates listed for $H$ and $K$ in Table 1 differ. 

	The brightnesses and colors of sources detected in $J, H,$ and $K$ over 
the $12 \times 12$ arcsec field are listed in Table 2; $K-L$ is also listed 
whenever measured. The random uncertainties in the photometry can be determined 
by comparing brightnesses of sources measured on independent frames. 
Such a comparison indicates that the uncertainties in $K$ are $\pm 0.02$ mag at 
$K = 10$, and $\pm 0.06$ mag at $K = 14$. 

	Table 2 also lists source positions as offsets, in arcsec, from IRS 7. 
The $x$ axis is aligned North-South; offsets to the North and/or East of IRS7 
are positive numbers. Sources with IRS numbers, identified from images 
published by Tollestrup, Capps, \& Becklin (1989) and DePoy \& Sharpe (1991), 
are indicated. The number of entries in Table 2 is mainly limited by the 
$J$ data, which suffers from relatively large extinction and 
poor image quality. Moreover, some of the brighter IRS sources, such as IRS 7, 
were saturated in $H$ and $K$, and hence are not listed in Table 2.

	We have compared the photometric measurements for the most luminous 
members of the IRS16 complex with those made by DePoy \& Sharpe (1991). The 
mean difference, $\Delta$, based on four IRS16 sources 
and in the sense DePoy \& Sharpe minus the values derived in the 
current study, is listed in Table 3 for each filter. 
The quoted uncertainties are the standard error of the 
mean. IRS16S was not considered for these comparisons, as it is located 
in an extremely crowded environment. The entries in Table 3 suggest that 
systematic differences, in the sense that the current 
measurements are the fainter of the two, may be present, 
although $\Delta$ differs from zero at less than the $2 - \sigma$ level 
in all four filters. Nevertheless, it is worth noting that values of $\Delta 
\leq 0$ are expected since the superior angular resolution of the 
current data acts to reduce contamination from nearby sources.

	While a number of other studies have published photometry of GC 
sources, the lower angular resolution of these data makes it difficult to 
obtain accurate measurements in crowded environments such as the IRS16 complex. 
The occultation measurements summarized by Simons, Hodapp, \& 
Becklin (1990) are an exception, and the $K$ brightnesses of 
the IRS16 components listed in Table 2 of that study agree 
within the estimated uncertainties with our values.

\section{Photometric Properties of GC Stars}

\subsection{Luminosity Functions and Sample Completeness}

	The luminosity function (LF) of all sources detected in the 
$K$ mosaic field is shown in Figure 4. The number counts decline when $K \geq 
16$, indicating that incompleteness becomes significant at 
this point, which corresponds roughly to $M_K = -1.5$ if $R_0 = 8$ kpc 
(Reid 1993) and A$_K = 3$ (DePoy \& Sharpe 1991) -- values for distance and 
reddening that will be adopted for the remainder of the paper. Sample 
completeness depends on crowding, which varies across the field; consequently, 
the brightness at which the sample is 100\% completeness will change with 
distance from SgrA*. The inflexion points in the $J$, $H$, and $L$ LFs indicate 
that these data are typically complete to $J = 18$, $H = 16.5$, and $L = 10$.

	Blum {\it et al.} (1996a) found that the $K$ LF of stars within an 
arcmin of the GC follows a power-law, with an exponent matching that in BW. The 
current data reach roughly 3 mag deeper than those discussed by 
Blum {\it et al.} (1996a), while probing an area much closer to SgrA*.
The dashed line in the lower panel of Figure 4 shows the 
power-law fit to the BW LF constructed by Tiede, Frogel, \& Whitford (1995), 
shifted along the vertical axis to match the GC data when $K \leq 16$. It is 
apparent that the $K$ LFs for BW and the GC have similar exponents. 
Therefore, the power-law exponent of the $K$ LF within a few 
arcsec of the GC is the same as its value at larger radii. 

	Tamblyn \& Rieke (1993) suggest that bright main sequence stars may be 
present near the GC, and that these objects could ionize the diffuse gas in 
this area. The most luminous Galactic main sequence O stars have M$_V = 
-5.5$ (Humphreys \& McElroy 1984) and $V-K \sim -1$ (Koornneef 1983), so that 
M$_K = -4.5$. With the adopted values for distance and reddening, 
these objects would have $K \geq 13$ near the GC. 
There is no evidence for a discontinuity in the LF when $K \geq 13$, 
suggesting that bright main sequence stars are not present in large numbers.

\subsection{Color-Magnitude and Two-Color Diagrams}

	The $(K, J-K)$, $(K, H-K)$, and $(K, K-L)$ color-magnitude diagrams 
(CMDs) are shown in Figures 5, 6, and 7. The data in the right hand panels of 
each figure have been corrected for distance and reddening using the values 
adopted earlier. The dashed lines in the left hand panels define the 100\% 
completeness limits; data falling above these curves have the smallest 
photometric errors, and hence form the basis for the photometric investigation 
discussed below.

	Spectra taken at $2\mu$m indicate that bright sources 
near the GC can be sorted into three different groups.
One group shows strong CO absorption (Lebofsky {\it et al.} 1982; Krabbe 
{\it et al.} 1995), and many of these stars may be evolving on the AGB (Blum 
{\it et al.} 1996b). Another group, which likely consists of evolved 
early-type stars, shows an absence of CO 
absorption (e.g. Rieke, Rieke, \& Paul 1989) and strong $2.058\mu$m He I 
emission (Krabbe {\it et al.} 1991, 1995; Libonate {\it et al.} 1995; Tamblyn 
{\it et al.} 1996). Finally, there are also heavily reddened objects that show 
a very steep, largely featureless continuum, which may be Young Stellar 
Objects (Krabbe {\it et al.} 1995); these sources tend to be too faint to be 
detected with the current data at $J$ and $H$. We have indicated on each CMD 
those sources in our field that have $2\mu$m spectra and fall into the first 
two groups, distinguishing between stars that show CO absorption and HeI 
emission.

	The spectroscopically identified early and late-type stars 
occupy very different positions on the $(K, J-K)$ CMD. 
The former fall along the dominant plume in the CMD, with $(J-K) \sim 
5.0 - 5.5$, while the latter fall along a redder sequence 
centered near $J-K = 6$, with a fainter peak 
brightness. Therefore, at least in the small field 
considered here, it appears that early and late-type stars can be separated 
on the $(K, J-K)$ CMD.

	The relatively red $(J-K)_0$ colors of early-type stars near the 
GC are somewhat surprising. McGregor, Hillier, \& Hyland (1988) obtained 
near-infrared spectra and photometry for a number of early-type supergiants 
in the Magellanic Clouds (hereafter MCs), and the six stars in their 
sample showing $2.058\mu$m HeI emission are plotted as 
filled squares in the right hand panel of Figure 5. These data were 
de-reddened using the $E(B-V)$ estimates given in Table 8 of McGregor et 
al. (1988), while an LMC distance modulus of 18.45 (van den Bergh 1989) was 
assumed. Two of the sources do not have 
independent color excesses, so the mean reddening of the other four 
sources was adopted for these objects.

	The MC He I emission stars have $(J-K)_0 \sim 0.4$, and fall well 
blueward of the GC early-type star sequence on the $(M_K, (J-K)_0)$ CMD. The 
$(J-H, H-K)$ two-color diagram, shown in Figure 8, 
reveals that while the early-type stars near the GC have 
$(J-H)_0$ colors similar to the MC objects, they have significantly redder 
$(H-K)_0$ colors. In fact, the $H-K$ colors of bright 
early-type stars near the GC are comparable to those of late-type stars. 
These comparisons suggest that the early-type stars near the GC have 
excess infrared emission with respect to similar objects in the MCs. 

	The $(K, K-L)$ CMD and the $(K-L, H-K)$ two-color diagram, plotted in 
Figure 9, further demonstrate the extent to which the photometric properties 
of bright early-type stars near the GC are affected by non-photospheric 
emission. In particular, the He I emission line sources have significantly 
redder $K-L$ colors than the CO absorption stars. 

	The stellar content of BW has been the subject of a number of 
investigations, and it is of interest to compare the photometric properties 
of late-type stars in BW with those near the GC. This comparison is 
made in the right hand panel of Figure 5, where
data for M giants in BW listed by Frogel \& Whitford (1987, 
hereafter FW) are plotted. There is good agreement between the GC and BW 
sequences. It is apparent from Figure 8 that late-type stars in 
these two environments also occupy similar regions of the $(J-H, H-K)$ 
two-color diagram.

	The $(K, H-K)$ CMD extends roughly 2 mag 
fainter in $K$ than the $(K, J-K)$ CMD, mainly because 
image quality improves while line-of-sight extinction decreases towards longer 
wavelengths. Unfortunately, the $(K, H-K)$ CMD is of only limited use for 
stellar content studies near the GC, due to excess infrared 
emission in the spectrum of bright early-type stars (see above), 
coupled with the lower temperature sensitivity inherent to $H-K$. 
In fact, there is no clear distinction between the colors of  
HeI emission and CO absorption sources in Figure 6. Blum {\it et al.} (1996a) 
attributed the inability to separate early and late-type stars on the $(K, 
H-K)$ CMD to differential reddening; however, it appears that 
excess infrared emission originating from early-type stars is 
also an important consideration.

	The scatter in the giant branch on the $(K, H-K)$ CMD 
is correlated with distance from SgrA*. To demonstrate 
this point, the field was divided into three equal area annuli, 
centered on SgrA*, and the resulting CMDs are compared in Figure 10.
There is a tendency for stars with $K \leq 10.5$ to be located in the 
inner annulus, a result that reflects the concentration of bright stars in the 
IRS16 complex. The comparatively narrow CMD in the vicinity of SgrA* is 
striking, and it is evident that the scatter along the color axis increases 
with radius. This trend suggests that differential reddening, the incidence of 
which will increase when data from areas with increasingly larger 
angular separations are combined, is significant in this field. 
Reddening variations of amplitude $\Delta A_V \sim 10 - 15$ mag would be 
capable of explaining the width of the CMD.

	If bright main sequence stars have only modest 
infrared excesses then it might be possible 
to separate these sources from the main body 
of more evolved objects on the $(K, H-K)$ CMD. The region on 
the $(K, H-K)$ CMDs that would be populated by early-type main sequence stars 
with no infrared excess is delineated by the dotted lines in Figure 10. The 
horizontal and vertical boundaries correspond to M$_K = -4.5$ and $(H-K)_0 = 
0$. It is perhaps not surprising that the largest number of blue 
objects are seen in the outermost annulus, 
where the photometry is deepest. We caution that these objects could be 
bulge stars subject to lower than average reddening, or foreground disk 
objects. Spectra will be required to determine the nature of these sources. 

\subsection{Comparisons with Isochrones}

	The $(K, J-K)$ CMD is compared with selected solar metallicity (z=0.02) 
and super metal-rich (z=0.05) red giant branch (RGB) and 
asymptotic giant branch (AGB) sequences from Bertelli {\it et 
al.} (1994) in Figure 11. Bertelli {\it et al.} (1994) list $M_V$ 
and $V-K$ for key evolutionary stages, such as the base and tip of the RGB. 
The solid lines in Figure 11 join the RGB-base and RGB-tip, while 
the dotted lines connect the RGB and AGB tips.
$J-K$ colors were computed from the published $V-K$ values using the relation 
defined by the BW M giant observations given by FW. 

	The metal-rich log(t) = 10.0 isochrones provide a good 
match to the giant branch, as defined by sources 
showing CO absorption. While this agreement is 
gratifying, caution should be exercised when assessing the significance of this 
result, as the model colors are sensitive to poorly 
determined quantities such as the mixing length, the relation 
between effective temperature and color, and chemical mixture. However, 
the color differences between the various isochrones are likely more 
reliable, and the comparisons in Figure 11 suggest that the scatter in the 
giant branch supports an age dispersion 
$\Delta$log(t) $\leq 1$ dex among the older stars near the GC.

	Luminosities predicted from evolutionary models are more reliable than 
colors. The brightest red stars in our field have $M_K \sim -7$ and the 
isochrones in Figure 11 suggest that these may be RGB-tip objects. The models 
also predict that if a large intermediate age population were present then 
red objects brighter than $M_K \sim -8$ should be detected; while these are 
abscent in the current data, a population of CO absorption 
sources with $M_K \sim -8$ is seen at larger distances from SgrA* (Blum {\it et 
al.} 1996b).

\section{Aperture Measurements of SgrA*}

	Eckart {\it et al.} (1995) resolved the region within an 
arcsec of SgrA* into a compact cluster of sources, none of 
which appear to be the optical counterpart to SgrA*. The $K$ brightnesses and 
$H-K$ colors of the stars in this complex that are resolved with the current 
data are summarized in Table 4, where $r$ is the angular separation from SgrA* 
in arcsec. The photometric properties of these stars are not 
significantly different from those of objects at larger radii, suggesting 
that the close proximity to SgrA* has not profoundly affected the 
near-infrared spectral energy distributions of these sources.

	Aperture measurements can be used to probe the unresolved stellar 
content near SgrA*. Eckart {\it et al.} (1993) and Close et al. (1995) 
concluded that the region within 0.5 arcsec of SgrA* is relatively blue, 
although this result was based on images having poorer angular resolution than 
that achieved here. In an effort to verify these earlier results, aperture 
measurements centered on SgrA* were made using the deep co-added images 
described in \S 2. These measurements were made using the PHOT routine in 
DAOPHOT, with the background measured in an annulus between 1 and 1.5 arcsec 
from SgrA*. In addition to sky emission, the background region also contains 
light from the surrounding bulge and the disk, so the aperture measurements 
presented below should be largely free of contamination from these components. 

	The $J-H$ and $H-K$ colors of SgrA*, corrected for reddening, are 
summarized in the second and third columns of Table 5. Three aperture sizes 
were used, the smallest of which matches the width of the central peak in the 
$H$ and $K$ PSF, while the largest covers the angular extent 
of the SgrA* cluster, as estimated from the $K$ image. 
The estimated uncertainties in these colors are $\pm 0.05$ mag. 
A second set of measurements were made with the bright members of the IRS16 
complex, the PSFs of which extend into the SgrA* region, 
subtracted from the image. The results, listed in the last two columns of 
Table 5, indicate that the near-infrared colors are not significantly 
affected by these sources.

	While the colors in Table 5 change with aperture size, 
they are consistent with a predominantly stellar origin. 
This is demonstrated in Figure 12, where the aperture measurements 
obtained after the bright IRS16 sources were subtracted from the data
are compared with those of LMC and SMC clusters studied by Persson {\it et al.} 
(1983). Any comparisons between the colors of bulge sources and MC 
clusters should be viewed with caution given the substantial 
differences in chemical composition that exist between these two 
systems. Nevertheless, the SgrA* measurements fall only slightly below the MC 
sequence in the two-color diagram, in qualitative agreement with what would be 
expected given the difference in metallicities between the GC and MCs.

\section{Discussion and Summary}

	High angular-resolution images obtained with the CFHT AOB have been 
used to investigate the stellar content in a $12 \times 12$ arcsec field 
centered on SgrA*. The $(K, J-K)$ CMD contains two distinct sequences that 
are populated by hot evolved massive stars and red giants. 
The hot stellar sequence is the most richly populated 
when $K \leq 12$, and has $J-K$ colors that are systematically redder than 
those of stars in the MCs that share similar spectroscopic characteristics. The 
$(J-H, H-K)$ and $(H-K, K-L)$ two color diagrams suggest that the 
difference between GC and MC sources is due to excess infrared emission from 
the GC population. This non-photospheric emission complicates population 
studies near the GC based on photometric data obtained longward of $1.5\mu$m, 
and explains, at least in part, why early and late-type GC stars occupy similar 
regions of the $(K, H-K)$ CMD. That early-type stars near the GC are more 
heavily obscured by circumstellar material than similar objects in the 
MCs is not an unexpected result if mass loss scales with metallicity, 
as predicted by radiative-driven wind theory (e.g. Kudritzki, 
Pauldrach, \& Puls 1987). In addition, higher metallicities in ejected material 
will lead to a greater circumstellar dust content.

	The giant branch stars within a few arcsec of SgrA* have photometric 
properties that are similar to those of giants in BW. The color and peak 
brightness of the giant branch locus on the $(K, J-K)$ CMD is well matched by 
metal-rich 10 Gyr isochrones, and the width of this sequence suggests that 
giants within a few arcsec of SgrA* span a 
range of ages $\Delta$log(t) $\leq 1$ dex. Therefore, if the 
Galactic bulge contains a large underlying population with an age in excess 
of 10 Gyr, as predicted from photometric (Holtzmann {\it et al.} 1993) and 
spectroscopic (Idiart {\it et al.} 1996) studies of BW, then the presence of a 
large intermediate age population within a few arcsec of the GC cannot be 
ruled out. Blum {\it et al.} (1996b) have identified a number of luminous 
AGB stars with $M_K \geq -8$ within a few arcsec of SgrA*. The 
Bertelli {\it et al.} (1994) isochrones, shown in Figure 11, suggest that stars 
of this brightness need not belong to a relatively young population; 
rather, they could be AGB stars with ages approaching 10 Gyr.

	Haller {\it et al.} (1996) and Sellgren {\it et al.} (1990) 
find a deficiency in the strength of CO absorption near SgrA*. 
While the $(K, J-K)$ CMD, which is complete to $K = 12$, 
shows only a modest giant branch population, the $K$ LF of our field, 
which extends 4 mag fainter, can be represented by a 
power-law exponent similar to that in BW, suggesting that a 
well-populated giant branch occurs within a few arcsec of SgrA*. This lends 
support to the suggestion made by Eckart {\it et al.} (1995) that the CO 
deficiency is due to contamination from an early-type population. However, we 
caution that the bulge is a three-dimensional system viewed in projection, 
so that some of the stars with small angular offsets from the GC 
may actually be foreground or background objects, located at large distances 
from the GC. 

	Aperture measurements centered on the largely 
unresolved complex of stars immediately surrounding 
SgrA* suggest that the light is photospheric in origin. The 
broad-band colors of this cluster are also relatively blue, and the discovery 
of a number of faint ($K \geq 14$) blue stars at larger distances from 
SgrA* suggests that a moderately bright main sequence component 
may be present in the vicinity of the GC. Nevertheless, spectra will 
be required to determine if these are bona fide main sequence stars at the 
distance of the GC, intrinsically red objects that are subjected to lower than 
average line-of-sight extinction, or foreground disk objects.

\vspace{0.5cm}
	Sincere thanks are extended to Eric Becklin for giving permission 
to use the IRTF $L'$ data. Our gratitude is also extended to an anonymous 
referee, who made numerous suggestions that greatly improved the paper.

\pagebreak[4]
\begin{center}
TABLE 1: Observing Log
\end{center}

\begin{center}
\begin{tabular}{ccccc}
\hline\hline
Filter & Date & Integration Time & FWHM & 50\% EE \\
 & & (sec) & (arcsec) & (arcsec) \\
\hline
$J$ & UT June 22/23 1996 & $6 \times 120$ & 0.42 & 0.52 \\
$H$ & UT June 21/22 1996 & $6 \times 50$ & 0.20 & 0.40 \\
$K$ & UT June 21/22 1996 & $24 \times 50$ & 0.16 & 0.35 \\
\hline
\end{tabular}
\end{center}

\pagebreak[4]
\begin{center}
TABLE 2: $JHKL$ Data
\end{center}

\begin{center}
\begin{tabular}{llllllll}
\hline\hline
Reference & IRS \# & $\Delta$x & $\Delta$y & $K$ & $J-K$ & $H-K$ & $K-L$ \\
\# & & (arcsec) & (arcsec) & & & & \\
\hline
1 & &  --3.9 &   0.7 & 13.28 &    6.13 & 2.05 & -- \\
2 & 7SE &    2.1 & -1.1 &    11.69 &    6.06 & 2.52 & 1.28 \\
3 & &    5.3 & --2.6 &   12.05 &    5.30 & 2.23 & -- \\
4 & &    2.3 & --2.4 &   14.36 &    1.65 & 0.87 & -- \\
5 & &    0.4 & --3.3 &   12.65 &    5.31 & 1.91 & -- \\
6 & &    5.5 & --5.1 &   11.93 &    5.26 & 2.17 & 4.08 \\
7 & 16NE &    2.7 & --4.5 &    9.24 &    4.99 & 1.98 & 1.75 \\
8 & &    2.9 & --5.0 &   11.94 &    5.01 & 2.41 & 2.13 \\
9 & &    5.6 & --5.6 &   11.95 &    5.65 & 2.53 & -- \\
10 & 16CC &    1.9 & --5.1 &   10.50 &    5.19 & 2.23 & 1.62 \\
11 & &    0.1 & --4.8 &   14.24 &    2.01 & 0.78 & -- \\
12 & 29S & --1.9 & --4.5 &   11.31 &    5.92 & 2.41 & 1.17 \\
13 & 16C &    1.0 & --5.1 &    9.92 &    5.64 & 2.41 & -- \\
14 & &  --1.6 & --5.8 &   13.03 &    4.96 & 1.80 & 2.04 \\
15 & 16SW-W &    0.8 & --6.5 &   10.09 &    5.33 & 2.02 & 1.85 \\
16 & &    0.5 & --7.1 &   11.49 &    5.29 & 2.05 & 2.54 \\
17 & &    0.1 & --7.1 &   13.58 &    5.18 & 2.24 & -- \\
18 & &    1.0 & --7.4 &   10.81 &    5.33 & 2.09 & 1.95 \\
19 & &    --0.3 & --7.7 &   11.20 &    5.25 & 2.08 & -- \\
20 & 13 &  --3.1 & --7.1 &   10.67 &    5.36 & 1.87 & -- \\
21 & 33W &  --3.1 & --8.1 &   11.20 &    5.92 & 2.10 & -- \\
22 & 33 &    0.4 & --8.6 &   10.24 &    5.54 & 2.18 & -- \\
23 & &    4.2 & --9.6 &   13.46 &    4.31 & 1.62 & -- \\
24 & &    0.5 & --9.5 &   12.36 &    5.56 & 2.14 & -- \\
25 & &    2.9 & --10.4 &   12.68 &    5.11 & 2.10 & -- \\
26 & &    4.5 & --11.2 &   12.25 &    5.38 & 2.17 & -- \\
27 & &  --5.1 & --9.3 &   12.43 &    6.78 & 2.13 & -- \\
28 & &    2.5 & --11.2 &   12.25 &    5.77 & 2.32 & -- \\
29 & 20 &   --1.2 & --10.8 &   10.89 &    6.08 & 2.11 & -- \\
30 & &    --0.5 & --11.6 &   11.34 &    6.36 & 2.32 & -- \\
\hline
\end{tabular}
\end{center}

\pagebreak[4]
\begin{center}
TABLE 3: Comparisons with DePoy \& Sharpe (1991)
\end{center}

\begin{center}
\begin{tabular}{cc}
\hline\hline
Filter & $\Delta$ \\
\hline
$J$ & $-0.42 \pm 0.27$ \\
$H$ & $-0.25 \pm 0.18$ \\
$K$ & $-0.16 \pm 0.12$ \\
$L$ & $+0.03 \pm 0.19$ \\
\hline
\end{tabular}
\end{center}

\pagebreak[4]
\begin{center}
TABLE 4: $H$ and $K$ Measurements of Sources Near SgrA*
\end{center}

\begin{center}
\begin{tabular}{ccc}
\hline\hline
r & $K$ & $H-K$ \\
(arcsec) & & \\
\hline
0.18 & 14.00 & 1.87 \\
0.17 & 14.02 & 1.78 \\
0.16 & 14.26 & 1.67 \\
0.26 & 14.51 & 1.92 \\
\hline
\end{tabular}
\end{center}

\pagebreak[4]
\begin{center}
TABLE 5: SgrA* Aperture Measurements
\end{center}

\begin{center}
\begin{tabular}{lcccc}
\hline\hline
Radius & $(J-H)_0$ & $(H-K)_0$ & $(J-H)_0$ & $(H-K)_0$ \\
(arcsec) & & & (No IRS16) & (No IRS16) \\
\hline
0.085 & 0.31 & 0.10 & 0.46 & 0.17 \\
0.170 & 0.21 & 0.01 & 0.26 & 0.05 \\
0.340 & 0.03 & --0.09 & 0.02 & --0.08 \\
\hline
\end{tabular}
\end{center}

\pagebreak[4]
\begin{center}
REFERENCES
\end{center}
\parindent=0.0cm

Allen, D. A., Hyland, A. R., \& Hillier, D. J. 1990, MNRAS, 244, 706

Bertelli, G., Bressan, A., Chiosi, C., Fagotto, F., \& Nasi, E. 1994, A\&AS, 
\linebreak[4]\hspace*{1.0cm}106, 275

Bica, E., Barbuy, B., \& Ortolani, S. 1991, ApJ, 382, L15

Blanco, V. M., \& Terndrup, D. M. 1990, AJ, 98, 843

Blum, R. D., DePoy, D. L., \& Sellgren, K. 1995, ApJ, 441, 603

Blum, R. D., Sellgren, K., \& DePoy, D. L. 1996a, ApJ, 470, 864

Blum, R. D., Sellgren, K., \& DePoy, D. L. 1996b, AJ, 112, 1988

Casali, M., \& Hawarden, T. 1992, JCMT-UKIRT Newsletter, 4, 33

Close, L. M., McCarthy, D. W. Jr., \& Melia, F. 1995, ApJ, 439, 682

Davidge, T. J. 1997, AJ, 113, 985

Davidge, T. J., Rigaut, F., Doyon, R., \& Crampton, D. 1997, AJ, 113, 2094

DePoy, D. L., \& Sharpe, N. A. 1991, AJ, 101, 1324

Eckart, A., \& Genzel, R. 1996, Nature, 383, 415

Eckart, A., Genzel, R., Hofmann, R., Sams, B. J., \& Tacconi-Garman, L. E. 
\linebreak[4]\hspace*{1.0cm}1993, ApJ, 407, L77

Eckart, A., Genzel, R., Hofmann, R., Sams, B. J., \& Tacconi-Garman, L. E. 
\linebreak[4]\hspace*{1.0cm}1995, ApJ, 445, L23

Forrest, W. J., Pipher, J. L., \& Stein, W. A. 1986, ApJ, 301, L49

Fried, D. L. 1965, J Opt Soc America, 55, 1427

Frogel, J. A., \& Whitford, A. E. 1987, ApJ, 320, 199

Gendron, E., \& L\'{e}na, P. 1994, A\&A, 291, 337

Haller, J. W., Rieke, M. J., Rieke, G. H., Tamblyn, P., Close, L., \& 
\linebreak[4]\hspace*{1.0cm}Melia, F. 1996, ApJ, 456, 194

Hodapp, K-W, Rayner, J., \& Irwin, E. 1992, PASP, 104, 441

Holtzman, J. A. {\it et al.} 1993, AJ, 106, 1826

Humphreys, R. M., \& McElroy, D. B. 1984, ApJ, 284, 565

Idiart, T. P., de Freitas Pacheco, J. A., \& Costa, R. D. D. 1996, AJ, 111, 
\linebreak[4]\hspace*{1.0cm}1169

Koornneef, J. 1983, A\&A, 128, 84

Krabbe, A., Genzel, R., Drapatz, S., \& Rotaciuc, V. 1991, ApJ, 382, L19

Krabbe, A. {\it et al.} 1995, ApJ, 447, L95

Kudritzki, R. P., Pauldrach, A., \& Puls, J. 1987, A\&A, 173, 293

Lebofsky, M. J., Rieke, G. H., \& Tokunaga, A. T. 1982, ApJ, 263, 736

Libonate, S., Pipher, J. L., Forrest, W. J., \& Ashby, M. L. N. 1995, ApJ, 
\linebreak[4]\hspace*{1.0cm}439, 202

McGregor, P. J., Hillier, D. J., \& Hyland, A. R. 1988, ApJ, 334, 639

Morris, M. 1993, ApJ, 408, 496

Nadeau, D., Murphy, D. C., Doyon, R., \& Rowlands, N. 1994, PASP, 106, 
\linebreak[4]\hspace*{1.0cm}909

Persson, S. E., Aaronson. M., Cohen, J. G., Frogel, J. A., \& Matthews, K. 
\linebreak[4]\hspace*{1.0cm}1983, ApJ, 266, 105

Reid, M. 1993, ARA\&A, 31, 345

Rich, R. M., \& Mighell, K. J. 1995, ApJ, 439, 145

Rieke, G. H., Rieke, M. J., \& Paul, A. E. 1989, ApJ, 336, 752

Rigaut, F., Doyon, R., Davidge, T. J., Crampton, D., Rouan, D., \& Nadeau, 
\linebreak[4]\hspace*{1.0cm}D. 1997, A\&A, submitted

Rigaut, F. {\it et al.} 1997, in preparation

Sellgren, K., McGinn, M. T., Becklin, E. E., \& Hall, D. N. B. 1990, 
\linebreak[4]\hspace*{1.0cm}ApJ, 359, 112

Simons, D. A., \& Becklin, E. E. 1996, AJ, 111, 1908

Simons, D. A., Hodapp, K.-W., \& Becklin, E. E. 1990, ApJ, 360, 106

Stetson, P. B. 1987, PASP, 99, 191

Stetson, P. B., \& Harris, W. E. 1988, AJ, 96, 909

Tamblyn, P., \& Rieke, G. H. 1993, ApJ, 414, 573

Tamblyn, P., Rieke, G. H., Hanson, M. M., Close, L. M., McCarthy, D. W. 
\linebreak[4]\hspace*{1.0cm}Jr., \& Rieke, M. J. 1996, 456, 206

Tiede, G. P., Frogel, J. A., \& Whitford, A. E. 1995, AJ, 110, 2788

Tollestrup, E. V., Capps, R. W., \& Becklin, E. E. 1989, AJ, 98, 204

van den Bergh, S. 1975, ARAA, 13, 217

van den Bergh, S. 1989, A\&A Review, 1, 111

\clearpage

\begin{figure}
\plotone{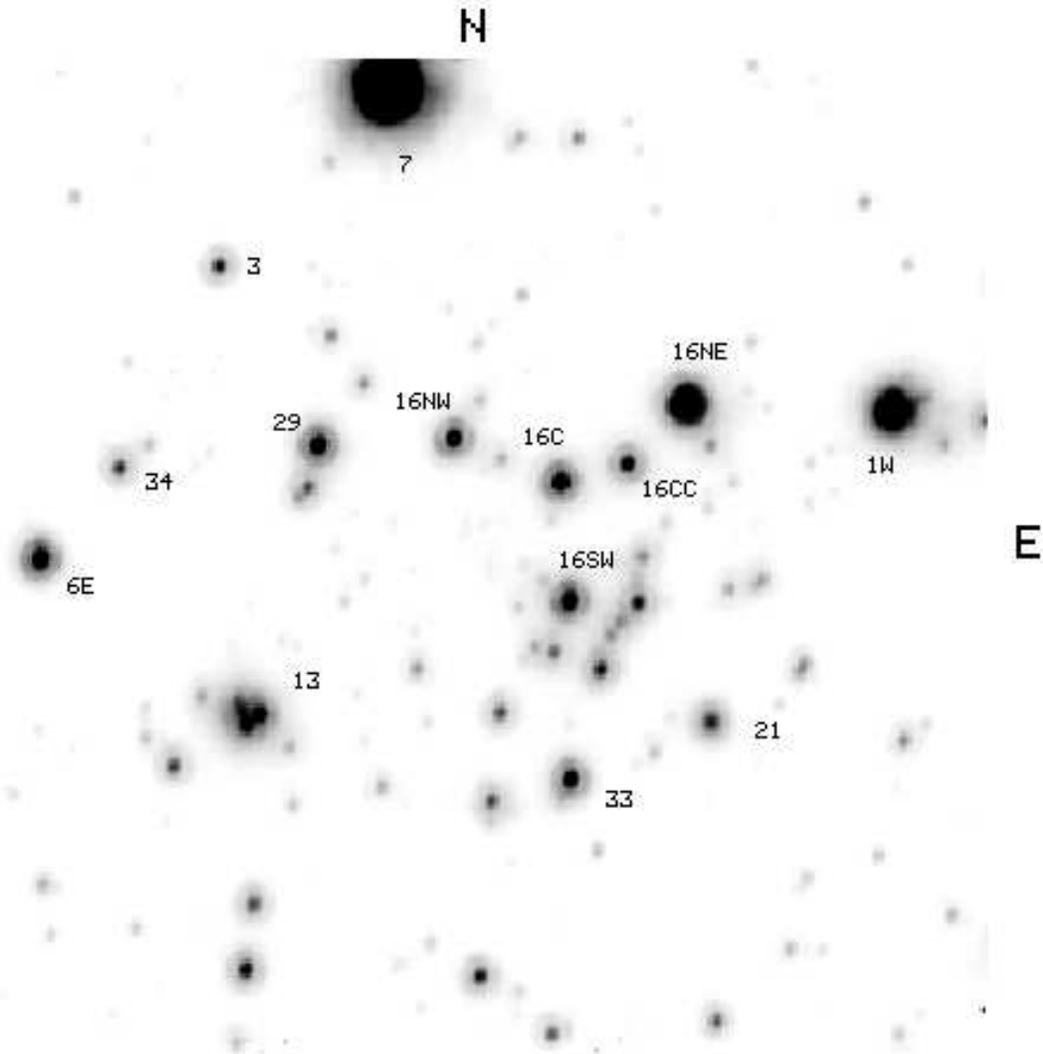}
\caption{The final $K$ mosaic of the GC field. The angular dimensions are 
$12.4 \times 11.9$ arcsec. The orientation of this image is such that a 
12.2 degree counter-clockwise rotation will position North at the top and East 
to the right. The Airey pattern can be seen around the brightest sources. SgrA* 
is at the approximate center of the image. Various IRS sources are labelled.}
\end{figure}

\begin{figure}
\plotone{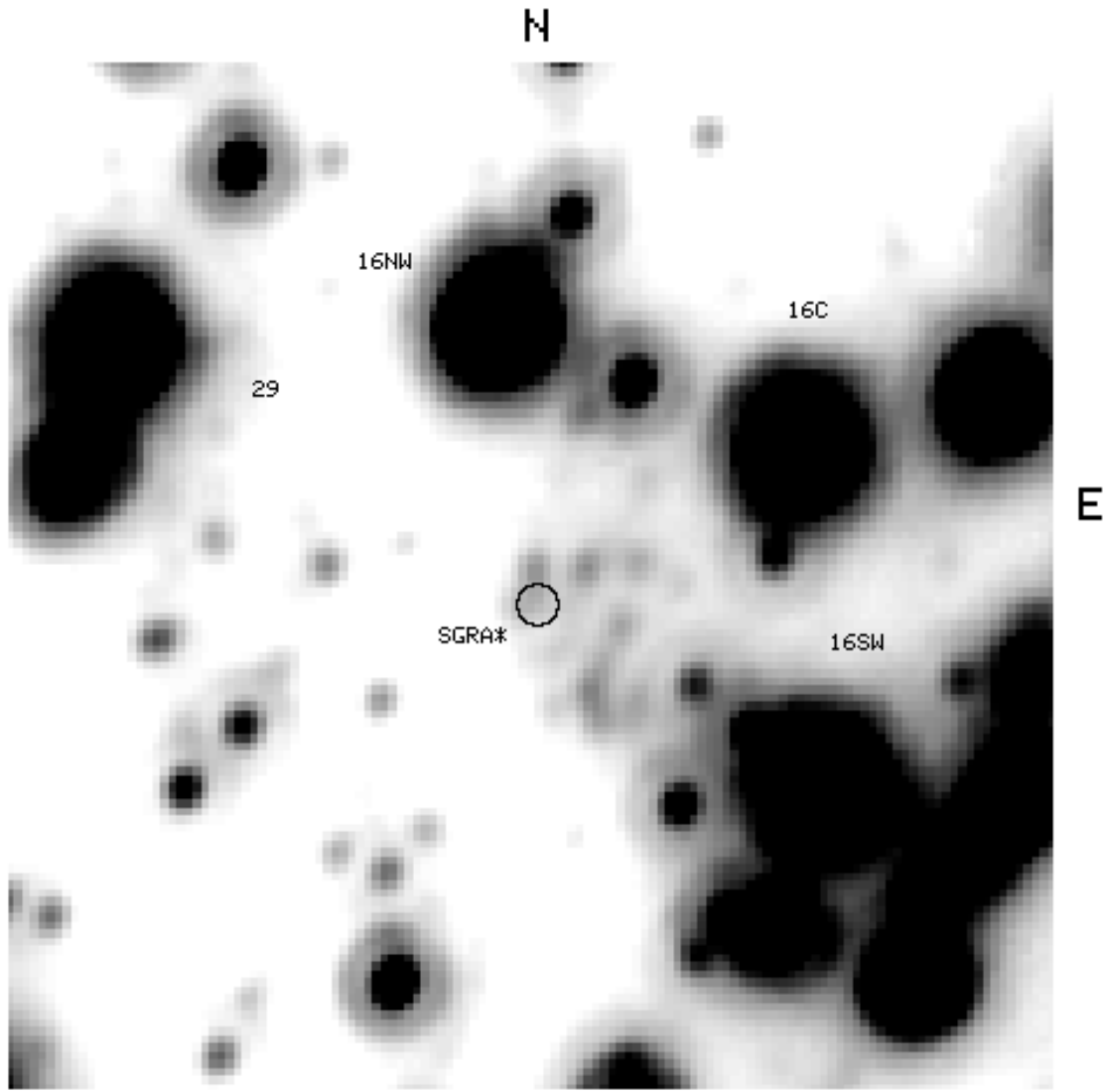}
\caption{The final deep $K$ image of the SgrA* field, which covers $4.7 \times 
4.7$ arcsec, and has an orientation identical to that of Figure 1. 
Selected IRS sources are identified, and the approximate location of SgrA* is 
indicated. The grey scale has been intentionally stretched to reveal faint 
sources.}
\end{figure}

\begin{figure}
\plotone{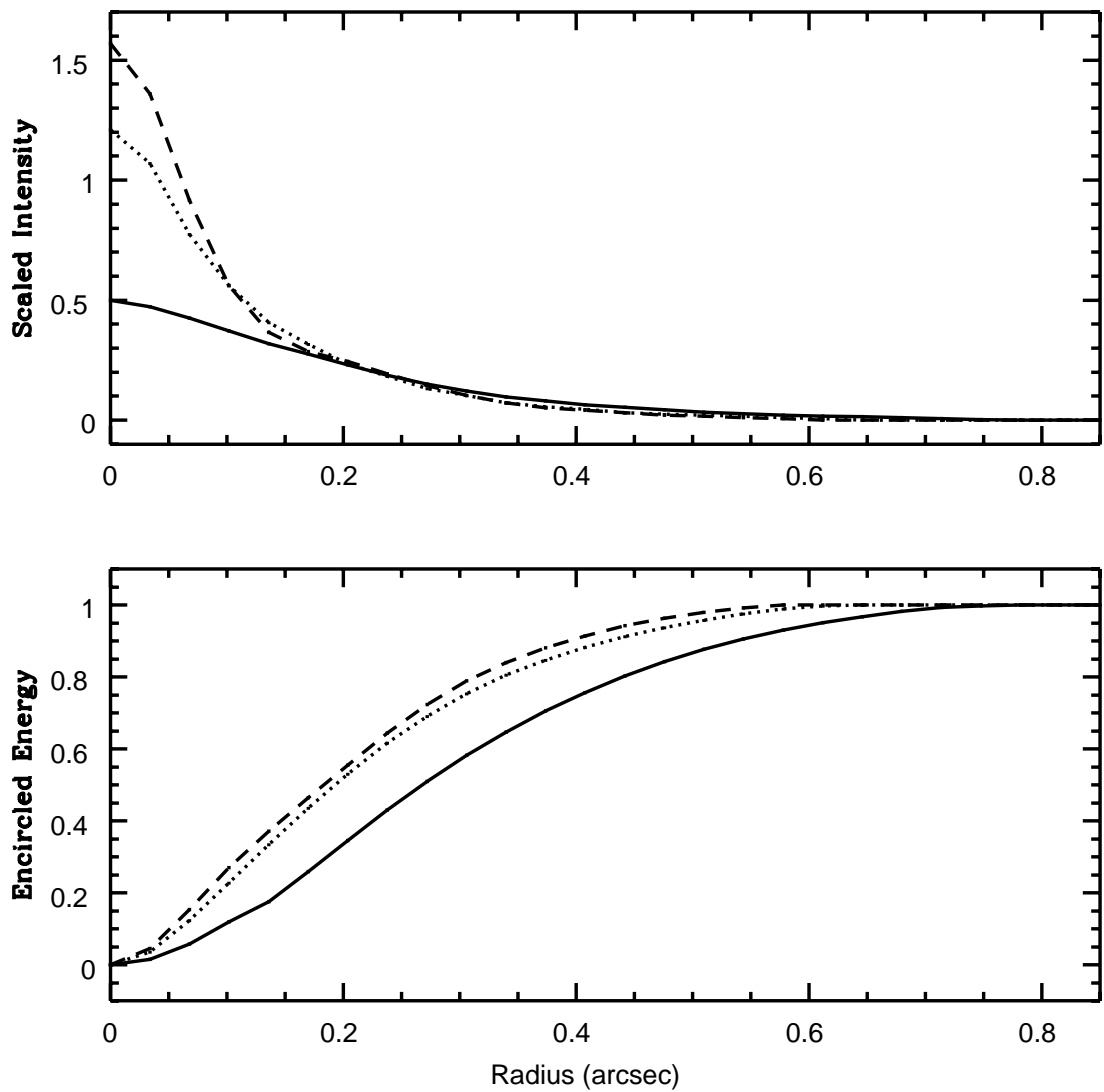}
\caption{Radially averaged intensity (top panel) and encircled energy 
(lower panel) curves derived from the $J$ (solid line), $H$ (dotted), and 
$K$ (dashed) PSFs. The intensity measurements have been scaled to a common 
integrated energy; consequently, the y-axis intercepts provide a 
relative measure of Strehl ratio. It is apparent from the lower panel that 
a significant amount of the total energy from a point source 
is located at radii in excess of 0.2 arcsec in all three filters.} 
\end{figure}

\begin{figure}
\plotone{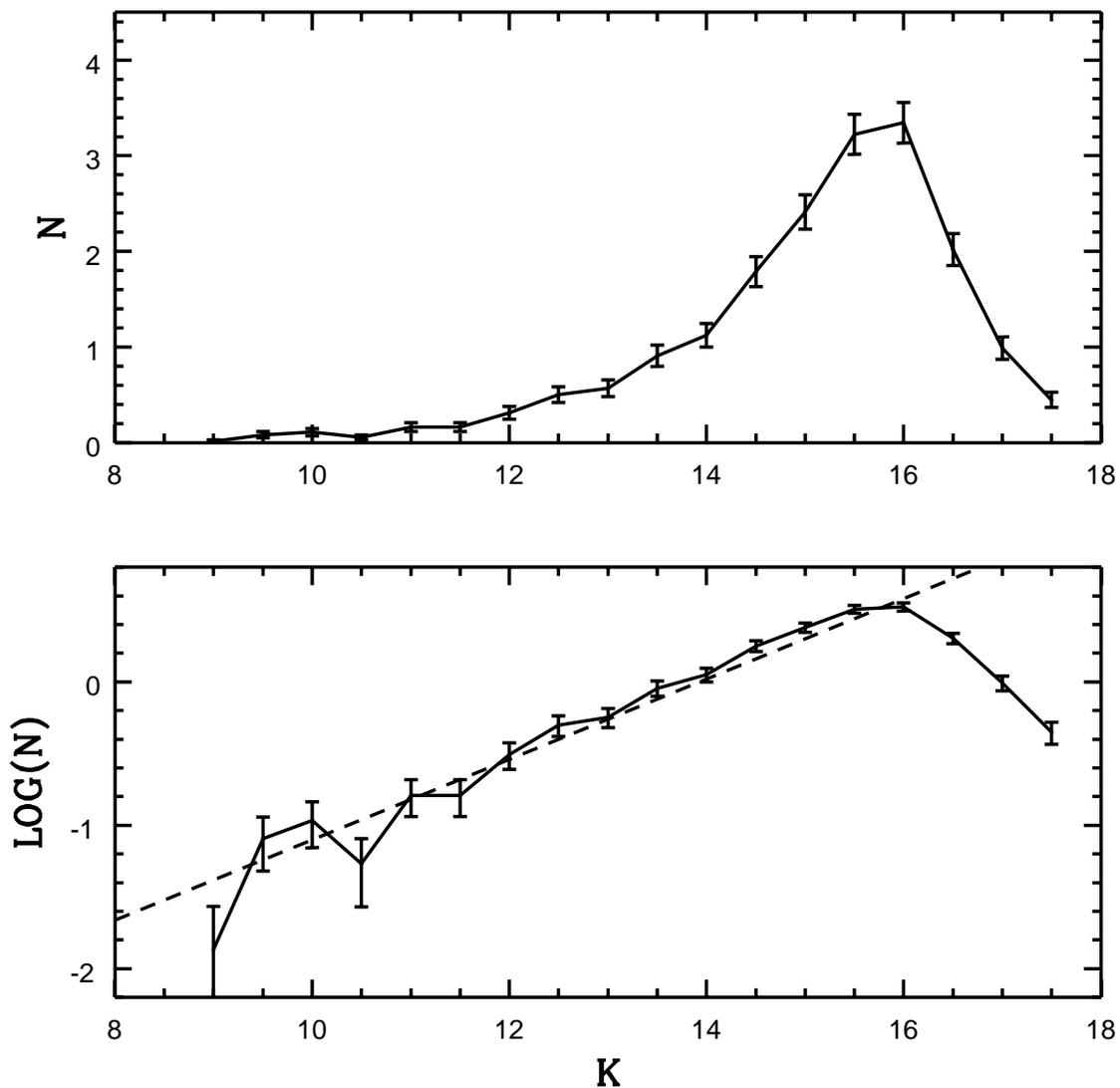}
\caption{The linear (top panel) and logarithmic (lower panel) luminosity 
functions of all sources detected in the $K$ mosaic 
field. $N$ is the number of stars per magnitude per square arcsec, and the 
errorbars are based on Poisson statistics. The dashed line in the lower panel 
shows the power-law exponent derived from the BW luminosity function of Tiede 
{\it et al.} (1995), shifted along the y axis to match the GC observations.}
\end{figure}

\begin{figure}
\plotone{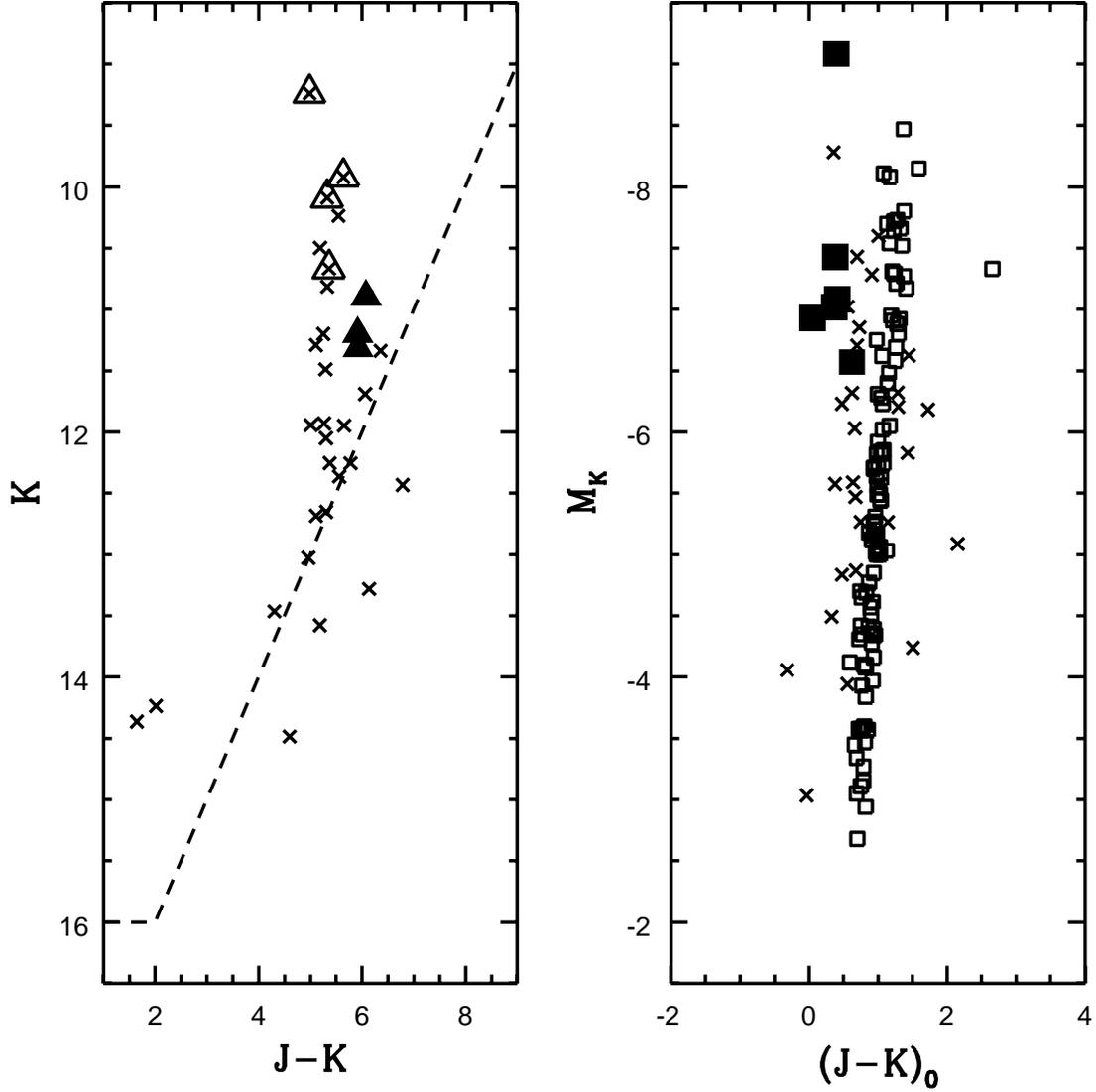}
\caption{The observed (left hand panel) and distance $+$ average reddening 
corrected (right hand panel) CMD for stars in the mosaic field. The 
data falling above the dashed line are 100\% complete. The filled triangles 
in the left hand panel indicate the sources IRS16NE, IRS16C, IRS16SW, 
and IRS13, which show $2\mu$m He I emission (e.g. 
Krabbe {\it et al.} 1995; Libonate {\it et al.} 1995). The filled 
triangles in the left hand panel indicate the sources IRS20, IRS29S, and 
IRS33W, which have $2\mu$m CO absorption, and hence are late-type stars (Krabbe 
{\it et al.} 1995; Blum {\it et al.} 1995b). The open squares in the right hand 
panel are data for M giants in BW from Table 1 of FW. Note that this sequence 
falls close to the giant branch defined by the stars 
showing CO absorption. The filled squares in the right 
hand panel are He I emission line stars studied by McGregor {\it et al.} 
(1988), assuming an LMC distance modulus of 18.45 (van den Bergh 1989).}
\end{figure}

\begin{figure}
\plotone{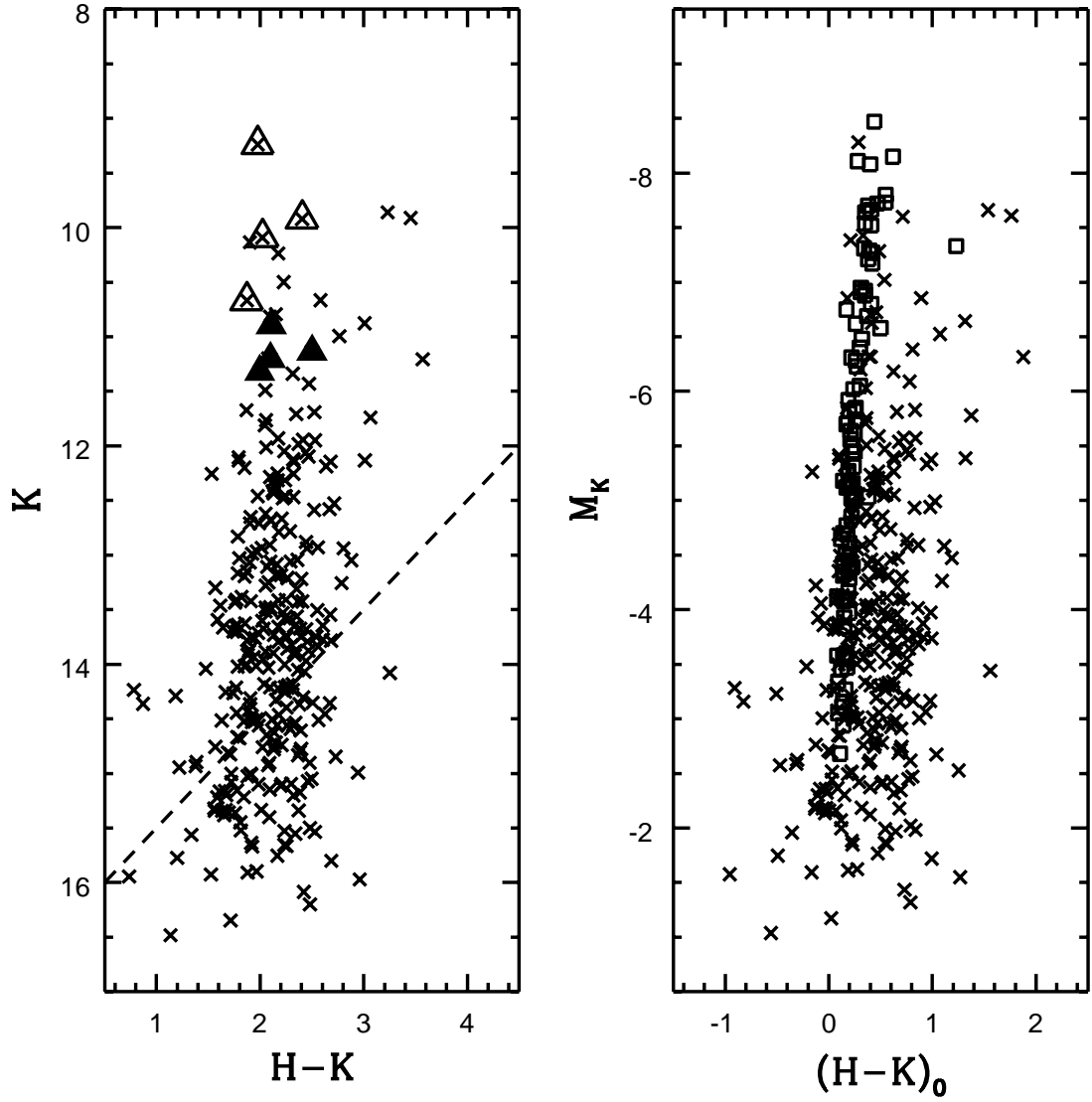}
\caption{Same as Figure 5, but for the $(K, H-K)$ CMD. Note the poor
color separation between the He I emission and CO absorption sources.}
\end{figure}

\begin{figure}
\plotone{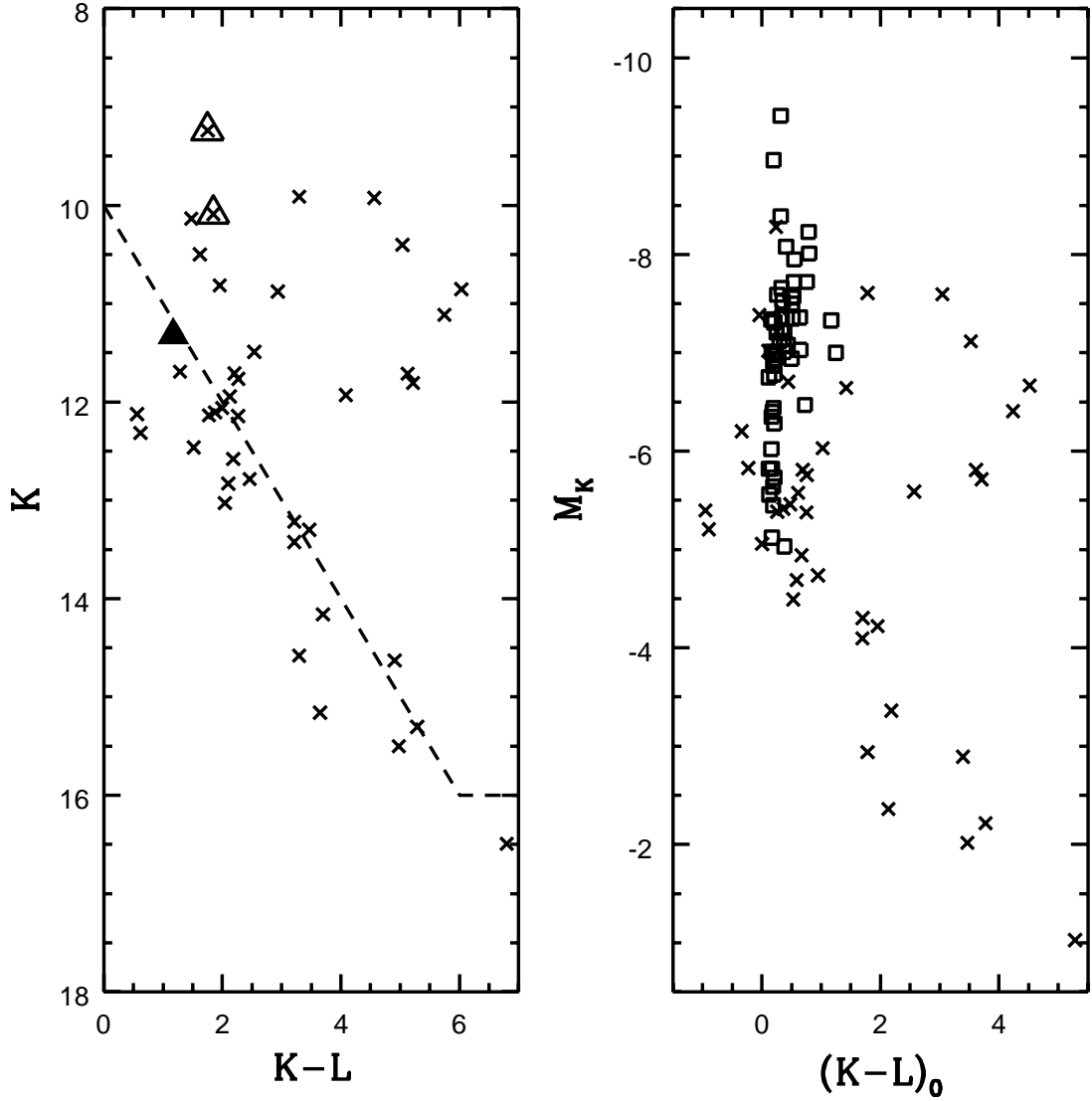}
\caption{Same as Figure 5, but for the $(K, K-L)$ CMD.} 
\end{figure}

\begin{figure}
\plotone{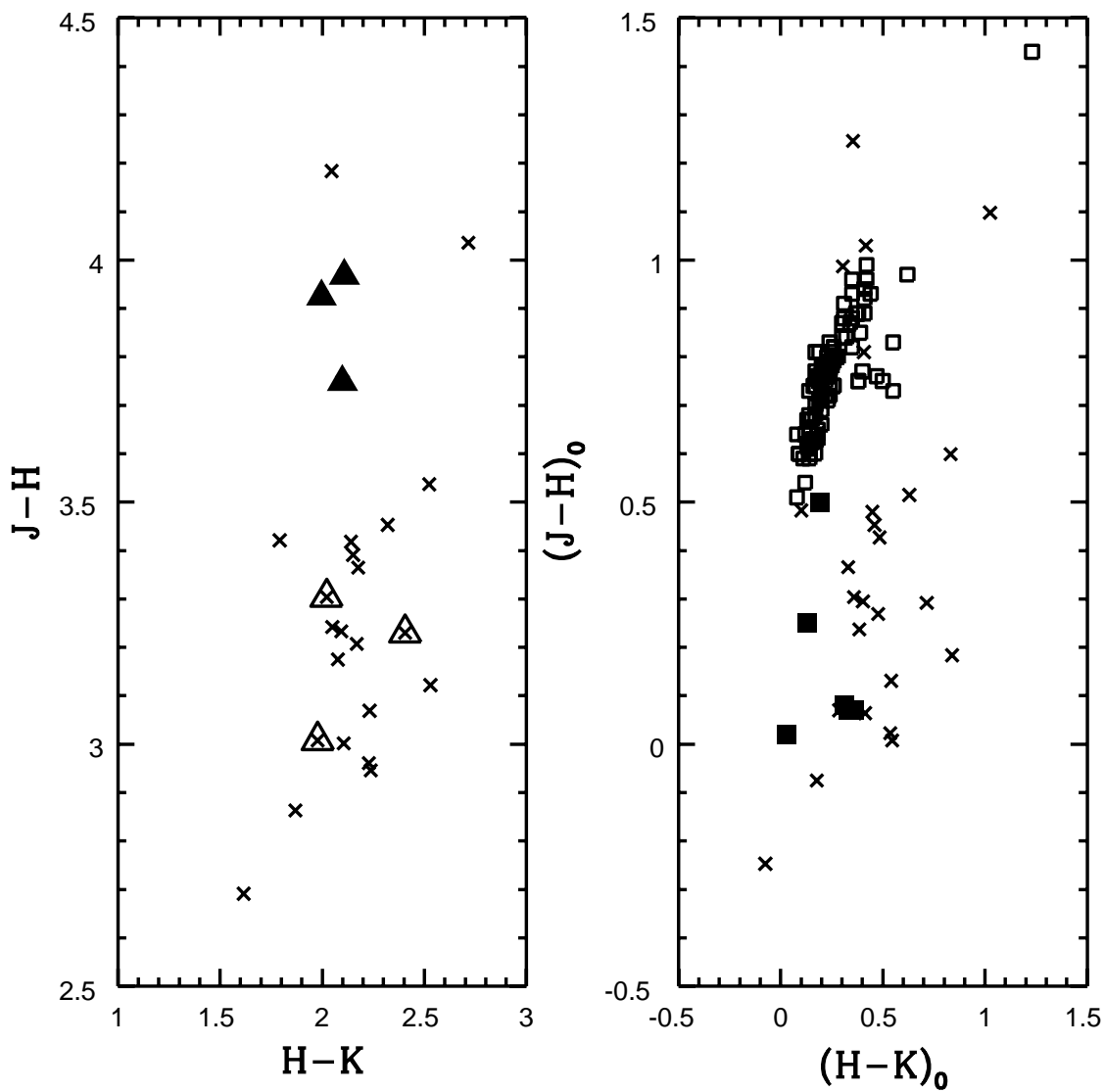}
\caption{The $(J-H, H-K)$ diagram for stars near the GC. The data in the 
right hand panel have been corrected for distance and reddening. The filled and 
open triangles in the left hand panel indicate stars with $2\mu$m spectra 
showing CO absorption and HeI emission, respectively. The open 
squares in the right hand panel are measurements for M giants in BW from FW. 
The filled squares in the right hand panel are data for Magellanic Cloud 
emission line stars, taken from McGregor {\it et al.} (1988).}
\end{figure}

\begin{figure}
\plotone{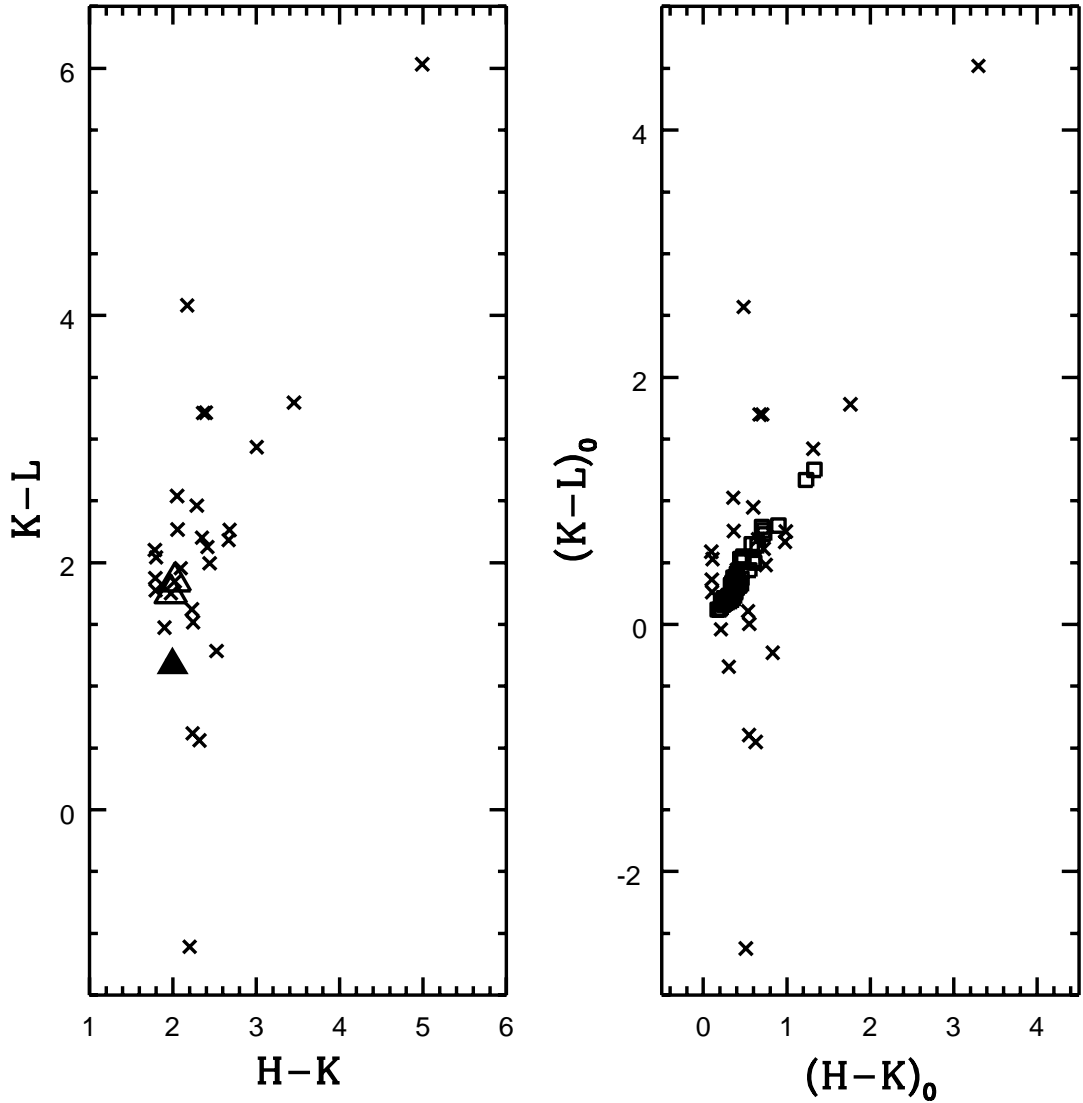}
\caption{The $(H-K, K-L)$ diagram for stars near the GC. The data in the right 
hand panel have been corrected for distance and reddening. Sources which, based 
on published $2\mu$m spectra, show line emission (open triangles) or CO 
absorption (filled triangles) are marked. The open squares in the right hand 
panel are measurements for M giants in BW from FW.}
\end{figure}

\begin{figure}
\plotone{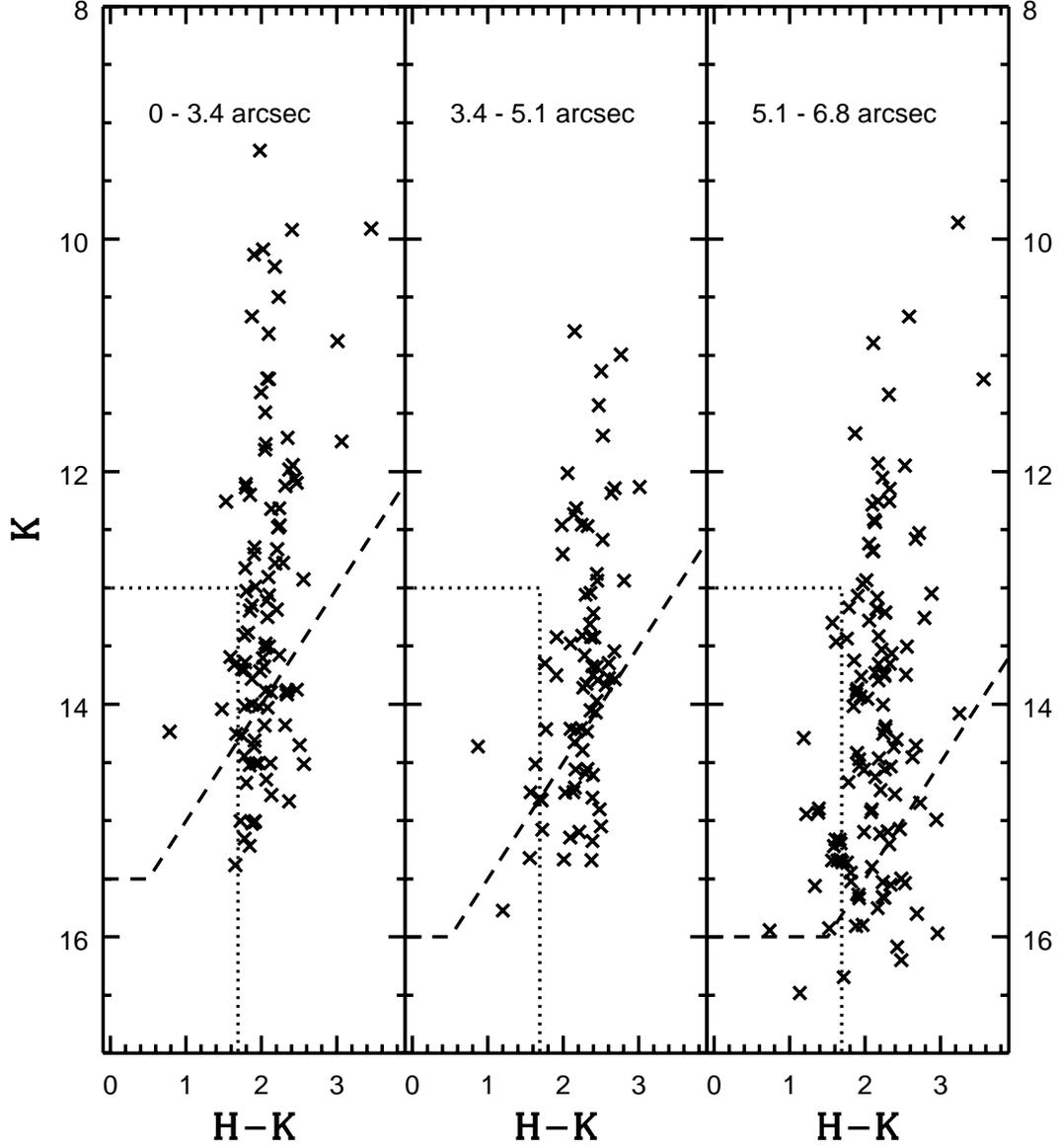}
\caption{The $(K, H-K)$ CMDs for three annuli. The intervals listed near the 
top of each panel are distances from SgrA*. Note the tendency for the giant 
branch to become progressively wider with increasing distance from SgrA*. The 
dashed lines define the 100\% completeness limits in each 
annulus. The dotted line defines the region with $K \geq 13$ and $H-K \leq 
1.69$, within which massive main sequence stars are expected to fall.}
\end{figure}

\begin{figure}
\plotone{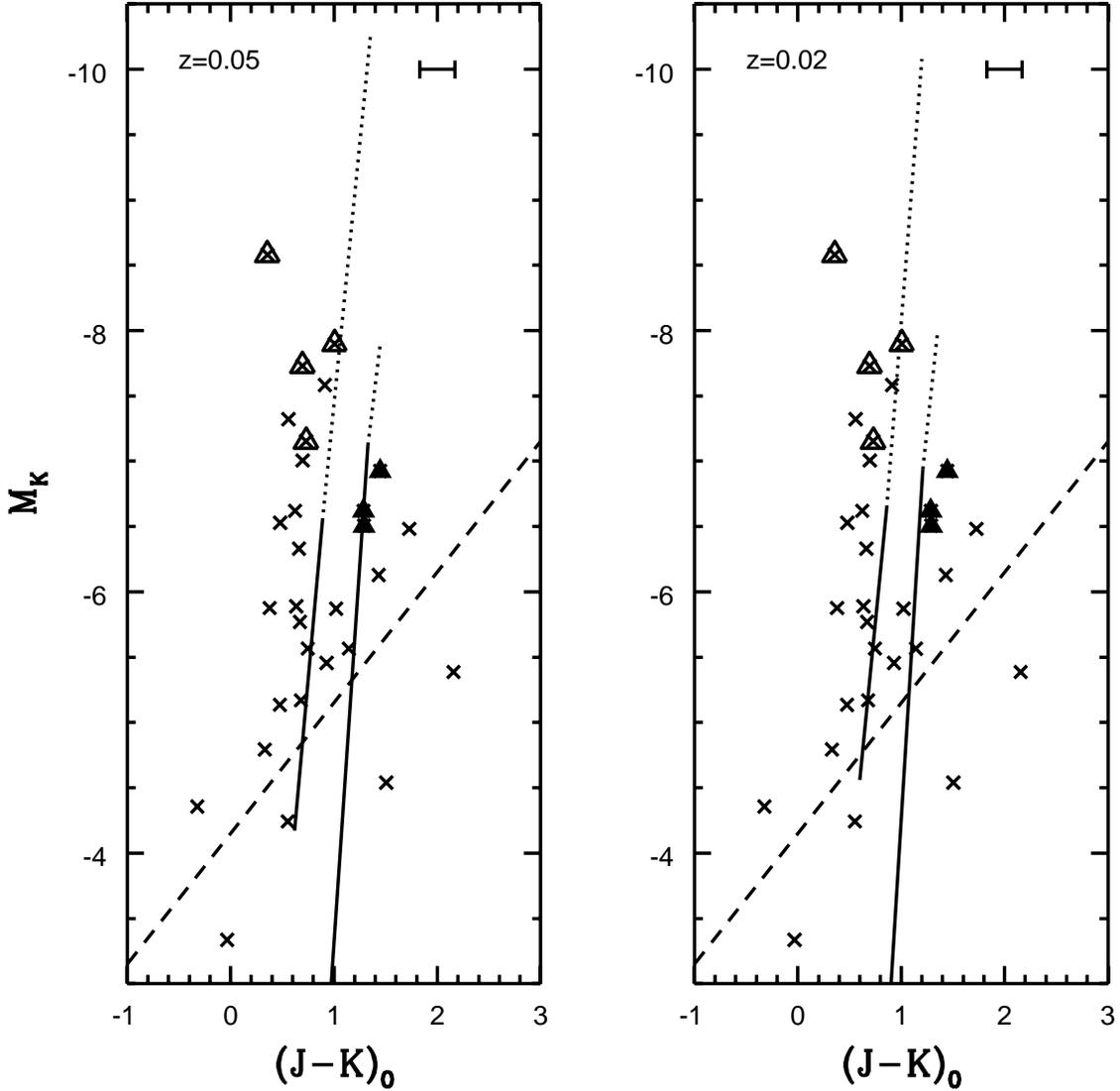}
\caption{The $(M_K, (J-K)_0)$ CMD of the GC field 
compared with theoretical isochrones from Bertelli {\it et al.} (1994). 
The dashed lines show the 100\% completeness limits, while the open and filled 
triangles indicate sources with $2\mu$m He I emission and CO absorption, 
respectively. The mean metallicities are listed in the top left hand corner 
of each panel. The solid lines connect the RGB-base with the RGB-tip, while the 
dotted lines join the RGB and AGB-tips. Sequences with 
log(t)=8.0 and log(t)=10.0 are shown in each panel. The errorbar shows 
the uncertainty in the adopted reddening value estimated from the 
scatter in the A$_V$ values in Table 1 of DePoy \& Sharpe (1991).} 
\end{figure}

\begin{figure}
\plotone{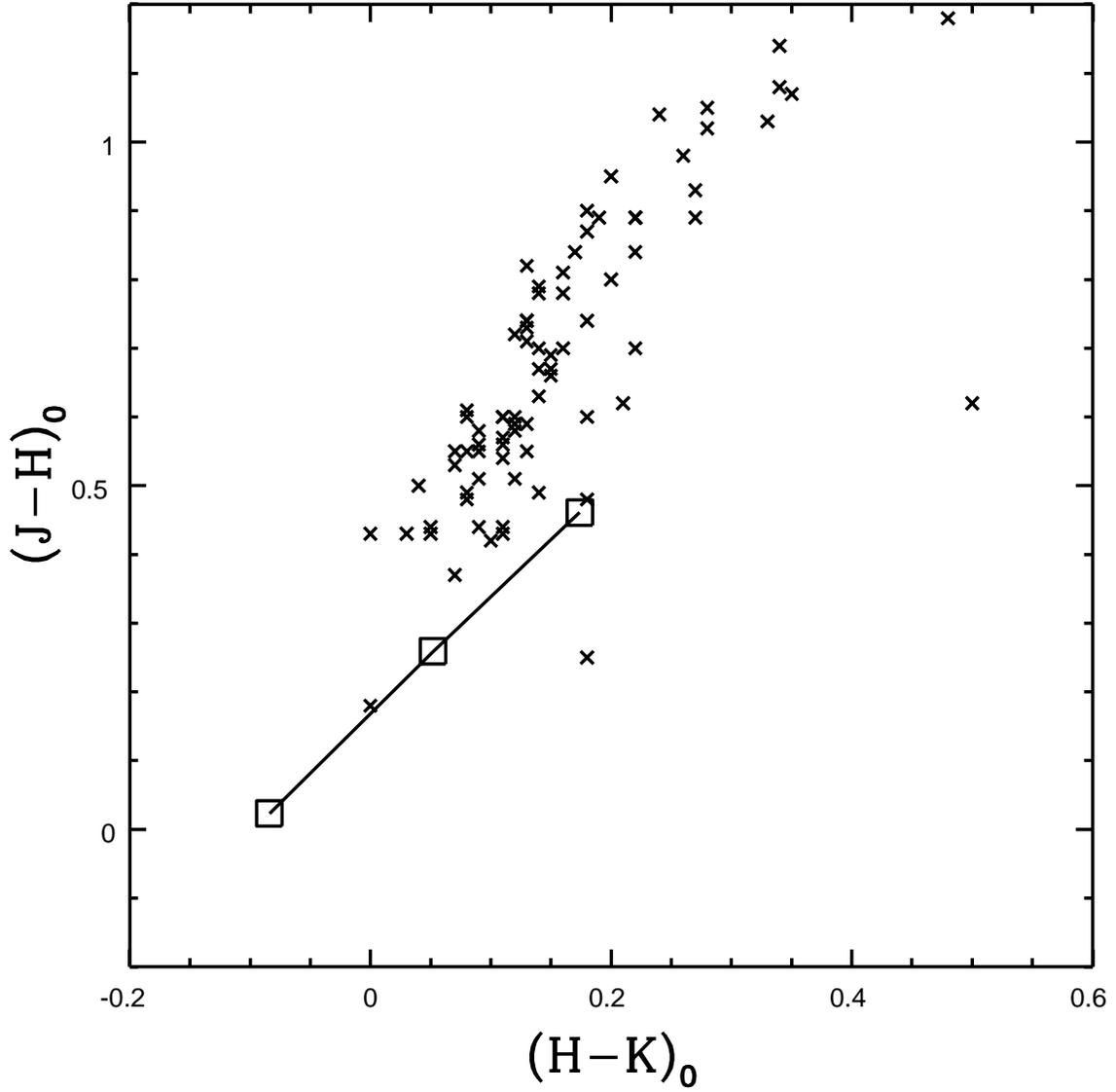}
\caption{The near-infrared two color diagram for aperture measurements of the 
star cluster surrounding SgrA* (open squares) and Magellanic Cloud clusters 
(crosses). The Magellanic Cloud measurements were taken from Persson {\it et 
al.} (1983).}
\end{figure}
\end{document}